\documentclass[a4paper,showpacs]{jpconf}
\usepackage{graphicx}

\usepackage{dcolumn}
\usepackage{bm}
\usepackage{amssymb}
\usepackage{float}
\usepackage[caption = false]{subfig}

\usepackage{footnote}

\usepackage{hyperref}
\usepackage{xcolor}

\begin{document}
\title{Status of CME Search Before Isobar Collisions and Methods of Blind Analysis From STAR}

\author{Prithwish Tribedy for the STAR collaboration}

\address{Physics Department, Brookhaven National Laboratory, Upton, NY}

\ead{ptribedy@bnl.gov}

\begin{abstract}
The STAR collaboration is currently pursuing the blind analysis of the data for isobar collisions that was performed at RHIC in the year 2018 to make a decisive test of the Chiral Magnetic Effect (CME)~\cite{starbur20}. Why is it so difficult to detect signals of CME in the experiment? Do we really understand different sources of background? Why observing similar charge separation between p/d+A and A+A does not stop us from pursuing the search for CME?  In this contribution, I attempt to address some of these questions and briefly outline a few recent STAR analyses based on new methods and observables to isolate the possible CME-driven signal and non-CME background contributions at the top RHIC energy. Finally, I describe the procedure for the blind analysis of the isobar data. An outstanding question remains -- what happens if we go down in energy?  I address this by discussing how the new event-plane detector (EPD) upgrade provides a new capability at STAR towards CME search using the data from the RHIC BES-II program. 

\end{abstract}

\section{Introduction}

 Finding a conclusive experimental evidence of the Chiral Magnetic Effect (CME) has become one of the major scientific goals of the heavy-ion physics program at the Relativistic Heavy Ion Collider (RHIC). 
 The existence of CME will be a leap towards an understanding of the QCD vacuum, establishing a picture of the formation of deconfined medium where chiral symmetry is restored and will also provide unique evidence of the strongest known electromagnetic fields created in relativistic heavy-ion  collisions~\cite{Kharzeev:1999cz,Kharzeev:2004ey}. 
 The impact of such a discovery goes beyond the community of heavy-ion  collisions and will possibly be a milestone in physics. Also, as it turns out, the remaining few years of RHIC run and analysis of already collected data probably provides the last chance for dedicated CME searches in heavy-ion  collisions in the foreseeable future.  
 Over the past years significant efforts from the STAR as well as other collaborations have been dedicated towards developing new methods and observables to isolate the possible CME-driven signal and non-CME background contributions in the measurements of charge separation across the reaction plane. 
The most widely studied experimental observable in this context is the $\gamma$-correlator, defined as $\langle\cos(\phi_a^{\alpha}+\phi_b^{\beta}-2\Psi_{RP}) \rangle$, where $\phi_a$ and $\phi_b$ denote the azimuthal angles of charged particles, $\alpha$ and $\beta$ are labels for the charge of the particles and $\Psi_{RP}$ is the reaction plane angle~\cite{Voloshin:2004vk}.  
The angle $\Psi_{RP}$ is expected to be strongly correlated to the direction of the magnetic field that enables the $\gamma$-correlator to be sensitive to signals of CME, more specifically, CME leads to a difference between same sign (SS,\, $\alpha=\beta$) and opposite sign (OS,\,$\alpha\ne\beta$) charge correlations: $\Delta\gamma=\gamma_{\rm OS} - \gamma_{\rm SS}$.  
The STAR time projection chamber (TPC) has a wide acceptance at mid-rapidity ($|\eta|\!<\!1$) that is used to detect $\phi_a$ and $\phi_b$. 
And, in STAR the proxy for $\Psi_{RP}$ can be played by: 1) second-order harmonic anisotropy plane $\Psi_2$ of produced particles at mid-rapidity measured by TPC, 2) the first-order plane due to the spectator neutrons ($\Psi_{\textsc{zdc}}$) detected by the zero degree calorimeters (ZDC), 3) the forward $\Psi_2$ plane using the STAR beam beam counter BBCs and 4) very recently using both the first and second-order harmonic anisotropy planes using the forward Event Plane Detector (EPDs). Each of these planes are expected to have more or less measurable correlations to B-field and serves their purpose for the CME search. 
The first measurement of non-zero $\Delta\gamma$ by the STAR collaboration goes back to ~\cite{Abelev:2009ac} where connections to several expectations from CME driven signals of charge separation was identified. Most importantly, the first measurement from STAR~\cite{Abelev:2009ac} also identified several possible contributions from non-CME effects in the experimental observation of non-zero $\Delta\gamma$. Several subsequent measurements from RHIC and LHC have confirmed this observation and provided many additional insights in that direction~\cite{Abelev:2009ac,Abelev:2009ad,Abelev:2012pa,Adamczyk:2013hsi, Adamczyk:2013kcb, Adamczyk:2014mzf, Adam:2015vje, Khachatryan:2016got, Acharya:2017fau,Sirunyan:2017quh,STAR:2019xzd}. In this contribution, I will focus only on RHIC results and refer to LHC results wherever necessary.

A major challenge that the  $\gamma$-correlator faces towards detecting signals of CME involves  large non-CME background sources that are: 1) correlated to $\Psi_{RP}$ and 2) independent of $\Psi_{RP}$. The distinction between the two sources must be carefully noted as they are crucial to the interpretation of several key measurements performed at both RHIC and LHC.

\section{Major challenges in isolating background}

\subsection{Background sources-I: reaction plane dependent correlations}

The possible background contamination due to the first source of $\Psi_{RP}$ dependent correlation was already alluded to in the reference where $\gamma$-correlator was first proposed~\cite{Voloshin:2004vk}. 
At that time only neutral resonance particles were identified as the major source of such background albeit thought to be sub-dominant. When a flowing neutral resonance decays it enhances the probability of a pair of opposite sign particles to move together along $\Psi_{RP}$. Such correlations 
lead to non-zero magnitudes of $\Delta\gamma$ mimicking CME. 
Later on, a more severe source of $\Psi_{RP}$ dependent  background due to correlated production of a pair of opposite charged particles due to local charge conservation (LCC) was proposed~\cite{Pratt:2010gy}. 
Parametrically, if $v_2$ is the elliptic flow and $N$ is the multiplicity the background contribution from resonance and LCC should go as $\Delta\gamma_{\rm bkg} \sim v_2/N$~\cite{Voloshin:2004vk} that is also verified by many model calculations~\cite{Schenke:2019ruo}.
Recently, many models that incorporate the same basic picture of particle production conserving charge locally from a flowing neutral matter, are able to very well explain measurements of $\Delta\gamma$ without invoking the physics of CME. 
Despite the success of background models experimental search of CME continued because of a number of reasons. Model predictions have large systematics since exact mechanism of hardronization is poorly understood, limited constraints from independent measurements are available. Above all, even the most state-of-the art background models fail to explain all qualitative features of the data (e.g. $\Delta\gamma$ in central collisions, see Fig.\ref{mixedharmonics}).  
While the models continue to refine their predictive power, over many years this largely lead to a major effort in beating the background sources in the measurement of charge separation along $\Psi_{RP}$. It is worth to mention that pheomenological predictions based on anomalous viscous hydrodynamics are now available that include both CME signal and background contribution and can be used to test the sensitivity of different observables~\cite{Jiang:2016wve}. 

\subsection{Background sources-II: reaction plane independent correlations}
The second major sources of non-CME background to $\Delta\gamma$ arises from reaction plane independent non-flow correlations. The possibility of such background was discussed in the first publication of charge separation from STAR~\cite{Abelev:2009ac}. One possible source of such background was identified to be three-particle correlations induced by mini-jet fragmentation which is known to: 1) influence the determination of event plane, 2) introduce more opposite charge correlation than same charge correlations. The combination of these two artifacts are supposed to lead to non-zero $\Delta\gamma$ and mimic CME signals. 
In Ref~\cite{Abelev:2009ac}, an indication of larger contribution of reaction plane independent background can already be seen in: 1) the sharp increasing strength of $\Delta\gamma$ towards peripheral events and, 2) large $\Delta\gamma$ in Cu+Cu than in Au+Au system at the same centrality. Both observations can be supported by \textsc{hijing} calculation. 
%
%

\section{Using small systems to estimate data driven background}
Small collision systems provide unique data-driven ways to measure charge separation in the background scenario. This is based on the idea that the direction of B-field is uncorrelated to the elliptic anisotropy plane of the produced particle with respect to which $\Delta\gamma$ is measured~\cite{Khachatryan:2016got,Belmont:2016oqp}. 
In low-multiplicity or min-bias collisions of small systems such planes are dominated by non-flow correlations from di-jets or momentum conservation. However, tell-tale signatures of collectivity have been observed in high multiplicity events of small collision systems -- the origin of which has been a widely discussed topic in our community. 
%
%
There are a few scenarios that decide whether the elliptic anisotropy plane measured in the experiment will be: 1) correlated to a geometric plane of participants if collectivity is due to hydrodynamics flow, 2) uncorrelated or less correlated to geometric plane if collectivity is due to non-hydrodynamic but other initial state momentum space correlations, e.g. from CGC or escape mechanism and, 3) dominated by non-flow from di-jets and momentum conservation if no collectivity is observed~\cite{Schenke:2019pmk}. Why is this important for CME search? It is important as these scenarios determine the nature of non-CME background that will dominate the measurements of $\Delta\gamma$ in small systems. It is also important to know what kind of baseline measurement do these small systems provide because our ultimate goal is to interpret measurements in heavy-ion collisions. For example, in the first scenario hydrodynamic flow driven background combined with local charge conservation will be the dominant source, important for heavy-ion  measurements in most centralities. For the second and third scenarios reaction plane independent background will be the dominant source, important for peripheral and smaller sized heavy-ion collisions. Nevertheless, the expectation is that CME signal in all such scenarios will be small as the B-field in small collision systems are weakly correlated to elliptic anisotropy plane other than some specific scenarios like what was discussed in Ref~\cite{Kharzeev:2017uym}. 
So in summary, small systems have the potential to provide baseline measurements for heavy-ion  collisions where CME signals are expected to disappear but different background sources will be present. 
The CMS measurement was the first to show that in overlapping multiplicity $\Delta\gamma$ measurements are quantitatively similar between p+Pb and Pb+Pb~\cite{Khachatryan:2016got}. STAR measurements performed in p+Au and d+Au systems show similar and in fact larger charge separation measured in terms of the scaled quantity $\Delta\gamma/v_{2} \times N_{\rm ch}$ than the same in Au+Au measurements~\cite{STAR:2019xzd}. Such observations are striking as they tell us that a very large value of $\Delta\gamma$ is expected even for $100\%$ background scenario. 

The following question is often asked. Does measurement in small systems completely rule out CME? Why do we still pursue the CME search?  
There are several reasons for not abandoning CME search in heavy-ion  collisions based on the observations from small collision systems. It is already known that $\Delta\gamma$ in heavy-ion  collisions suffer from major background, the possible existence of CME driven signal has become more of a quantitative question. Therefore only a quantitative baseline will serve our purpose. So a better question to ask is whether small system measurements can provide direct quantitative baseline for heavy-ions. 
Heavy-ion measurements for CME search are performed where the system size, multiplicity do not necessarily  overlap with that of small systems. It is not straightforward to extrapolate the quantitative background baselines for $\Delta\gamma$ into such unknown territories where change of physics is eminent. For example, $\Delta\gamma$ measured for $N_{\rm ch}=10$ in p/d+Au maybe a good baseline for A+A at the same multiplicity but may not serve as quantitative baselines for $\Delta\gamma$ in Au+Au at $N_{\rm ch}=100$. 
One may try to make a projection under some working assumptions but that will lead to a qualitative baseline and defeats the major purpose of using small systems as direct quantitative baselines. This is where isobar collisions come in -- that ensures measurements in two systems with very similar size and shape are compared. 
It is also difficult to conclude that the case of CME is ruled out entirely based on the raw $\Delta\gamma$ measurements between p/d+Au and Au+Au. In lieu of which several variants of $\Delta\gamma$, as well as alternative observable such as $R-$observable, signed balance function has been developed to quantify the signals of CME~\cite{Magdy:2017yje,Tang:2019pbl}. The measurements based on $R-$observable show qualitative difference in p/d+Au and Au+Au ~\cite{rcorrelator} -- that is discussed in the following section.  

\section{The way forward}

With the aforesaid introduction on the challenges to disentangle CME from non-CME background I would like to now proceed with the possible solutions to overcome such a problem. Many cleaver ideas have been proposed and applied to existing data. The general consensus is that measurement from the isobar collisions (Ru+Ru that has $10-18\%$ higher B-field than Zr+Zr) provides the best solution to this problem. In following sections of this conference proceedings I would like to mention a few such recent efforts such as: 1) Differential measurements of $\Delta\gamma$ to identify and quantify backgrounds, 2) measurement of higher order harmonics of $\gamma$-correlator, 3) exploiting the relative charge separation across participant and spectator planes, 4) the use of R-observable to measure charge separation and 5) the use of signed balance function. The first three approaches are based on aforementioned three-particle correlator and the last two employ slightly different approaches to quantify charge separation. 
There have been many more developments in the recent times and also many LHC measurements have been performed but I will specifically focus on these five approaches because they will be explored with the isobar data. The following five sections describe these procedures in brief with comments on the outlook for isobar blind analysis (see~\cite{starbur20} for more details). 

\section{Differential measurements of $\Delta\gamma$ to identify and quantify background}
\subsection{Invariant mass dependence of charge separation}
Differential measurements of $\Delta\gamma$ with invariant mass and relative pseudorapidity provide interesting prospects to identify and quantify the sources of flow and non-flow driven backgrounds. The idea to use invariant mass is simple and was first introduced in Ref~\cite{Zhao:2017nfq}. Resonances are widely identified by observing structures in the invariant mass spectra of the decay daughters. Take a pair of opposite sign pions for example, a large fraction of them come from the neutral resonances that show up in the invariant mass spectrum of $m_{inv}(\pi^+ + \pi^-)$. If we restrict the analysis to pairs of pions, differential measurements of $\Delta\gamma$ with $m_{inv}(\pi^+ + \pi^-)$ should also show similar peak like structures if background from neutral resonances dominate the charge separation. Indeed similar peak structures are observed and a careful analysis is performed by STAR collaboration to extract the possible fraction of CME signals from measurements~\cite{invmass}. This analysis relies on the assumption that CME signals do not show peak like structures in $m_{inv}(\pi^+ + \pi^-)$ therefore calls for more theoretical inputs in this direction. 

\subsection{Relative pseudorapidity dependence}
The relative pseudorapidity dependence of azimuthal correlations are widely studied to identify sources of long-range components that are dominated by early time dynamics as compared to late time correlations that are prevented by causality to appear as short-range correlations. The same can be extended to charge dependent correlations that provides the impetus to explore the dependence of $\Delta\gamma$ on the pseudorapidity gap between the charge carrying particles $\Delta\eta_{ab}=|\eta_a-\eta_b|$ in $\langle\cos(\phi_a^{\alpha}+\phi_b^{\beta}-2\Psi_{RP}) \rangle$. Such measurements have been performed in STAR with Au+Au and U+U data. It turns out that the possible sources of short-range correlations due to photon conversion of $e^+-e^-$, HBT and Coulomb effects can be identified and described as Gaussian peaks at small $\Delta\eta_{ab}$, the width and magnitude of which strongly depend on centrality and system size. Going to more peripheral centrality bins, it becomes harder and harder to identify such components as they overlap with sources of di-jets fragmentation that dominates both same-sign and opposite sign correlations. An effort to decompose different components of $\Delta\gamma$ via study of $\Delta\eta_{ab}$ can be challenging although a clear sign of different sources of correlations are visible in change of shape of individual same-sign and opposite sign measurements of $\gamma$-correlator~\cite{Tribedy:2017hwn}.  

In any case, these differential measurements of $\Delta\gamma$ in isobar collisions provide the prospect to extract the $m_{inv}(\pi^+ + \pi^-)$ and $\Delta\eta$ dependence of CME signals that will provide much deeper insights on the origin of the effect. 

%

\section{Mixed harmonics measurements with second and third order event planes}
\begin{figure}
\includegraphics[width=1.\textwidth]{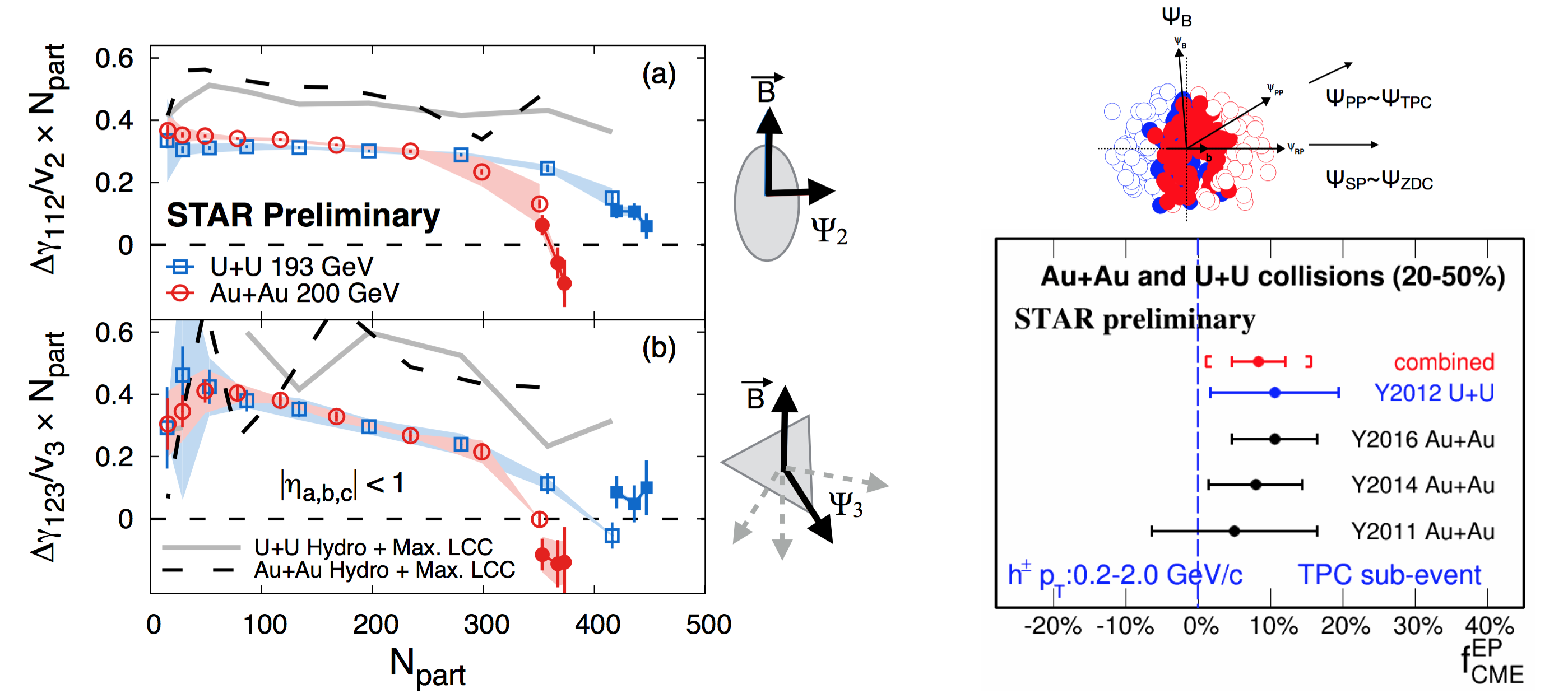}
\caption{\label{mixedharmonics} (Left) Measurement of charge separation along second and third order event planes in Au+Au and U+U collisions. (Right) Fraction of possible CME signal in the measurement of $\Delta\gamma$ with respect to spectator and participant planes~\cite{Zhao:2020utk}.}
\end{figure}

In order to proceed in this section it is better to rewrite the 
conventional $\gamma$-correlator by a more general notation as $\gamma_{112}=\langle\cos(\phi_a^{\alpha}+\phi_b^{\beta}-2\Psi_{2})\rangle$. The idea is to measure charge separations across the third harmonic event plane by constructing a new correlator $\Delta\gamma_{123}=\gamma_{123}(OS)-\gamma_{123}(SS)$, where $\gamma_{123}=\langle\cos(\phi_a^{\alpha}+2\phi_b^{\beta}-3\Psi_{3})\rangle$ that was introduced by CMS collaboration in Ref~\cite{Sirunyan:2017quh}. Since the $\Psi_3$ plane is random and not correlated to B-field direction (see Fig.\ref{mixedharmonics}), $\gamma_{123}$ is purely driven by non-CME background, the contribution of which should go as $v_3/N$. This is very useful to contrast signal and background scenario by comparing the measurements in two isobaric collision systems. Since Ru+Ru has larger B-field than Zr+Zr but have comparable background, the case for CME would be as follows: $(\Delta\gamma_{112}/v_2)^{\rm Ru+Ru}/(\Delta\gamma_{112}/v_2)^{\rm Zr+Zr}>1$ and $(\Delta\gamma_{112}/v_2)^{\rm Ru+Ru}/(\Delta\gamma_{112}/v_2)^{\rm Zr+Zr}>(\Delta\gamma_{123}/v_3)^{\rm Ru+Ru}/(\Delta\gamma_{123}/v_3)^{\rm Zr+Zr}$. Fig.\ref{mixedharmonics} (left) shows the measurement of these observables in U+U and Au+Au collisions. Within the uncertainties of the measurements, no significant difference in the trend of $\Delta\gamma_{112}/v_2$ and $\Delta\gamma_{123}/v_3$ is observed for the two collision systems except for the very central events. Predictions from hydrodynamic model calculations with maximum possible strength of local charge conservation~\cite{Schenke:2019ruo} is shown on the same plot. Overall observation indicates background dominate the measurements and a similar analysis of the isobar data is highly anticipated.


\section{Charge separation along participant and spectator planes}
This analysis makes use of the fact that B-field driven signal is more correlated to spectator plane in contrast to flow-driven background which is maximum along the participant planes. 
The idea was first introduced in Ref~\cite{Xu:2017qfs} and later on followed up in Ref~\cite{Voloshin:2018qsm}. It requires measurement of $\Delta\gamma$ with respect to the plane of produced particles, a proxy for participant plane as well as with respect to the plane of spectators. In STAR the two can be done by using $\Psi_2$ from TPC and $\Psi_1$ from ZDC respectively. 
The approach is based on three main assumptions: 1) measured $\Delta\gamma$ has contribution from signal and background that can be expressed as $\Delta\gamma=\Delta\gamma^{\rm bkg}+\Delta\gamma^{\rm sig}$, 2) the background contribution to $\Delta\gamma$ should follow the scaling $\Delta\gamma^{\rm bkg}(\textsc{tpc})/\Delta\gamma^{\rm bkg}(\textsc{zdc})=v_{2}(\textsc{tpc})/v_{2}(\textsc{zdc})$ and, 3) the signal contribution to $\Delta\gamma$ should follow the scaling $\Delta\gamma^{\rm sig}(\textsc{tpc})/\Delta\gamma^{\rm sig}(\textsc{zdc})=v_{2}(\textsc{zdc})/v_{2}(\textsc{tpc})$. The first two  have been known to be working assumptions, widely used for a long time and can be used to test the case of CME~\cite{Voloshin:2018qsm} if $\left(\Delta\gamma/v_2\right)(\textsc{zdc})/\left(\Delta\gamma/v_2\right)(\textsc{tpc})>1$. The validity of the last one was studied and demonstrated in Ref~\cite{Xu:2017qfs}. Using all three equations one can extract~\cite{Zhao:2020utk} the fraction of possible CME signal $f_{\textsc{cme}}=\Delta\gamma^{\rm sig}/\Delta\gamma$ in a fully data-driven way as shown in Fig.\ref{mixedharmonics}(right). This analysis will be done with the isobar data and the case for CME will be $f_{\textsc{cme}}^{\rm Ru+Ru}>f_{\textsc{cme}}^{\rm Zr+Zr}>0$. 

\section{Alternate measure: The novel R-observable}
\begin{figure}
\includegraphics[width=1.\textwidth]{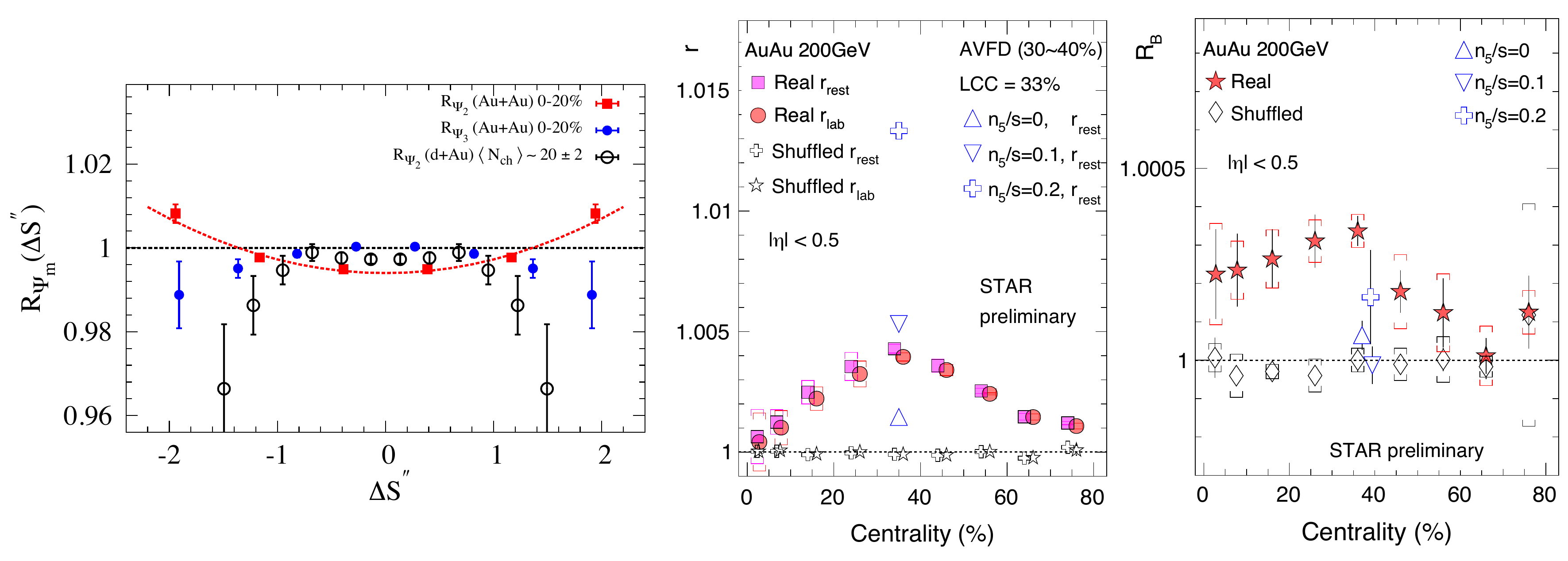}
\caption{\label{rcorrelator} (Left) The R-observable shown for different collision systems, concave shape is consistent with CME expectation~\cite{rcorrelator}. (Right) The two main quantities $r$ and $R_B$ derived from signed balance function, deviation from unity is consistent with CME expectation~\cite{Lin:2020jcp}.}
\end{figure}
The $R$-observable is actually a distribution, introduced in Ref~\cite{Magdy:2017yje}, and defined as the ratio of two distribution functions of the quantity $\Delta S$ parallel and perpendicular to B-field direction defined as $R_{\Psi_m} (\Delta S)=C_{\Psi_m}(\Delta S)/C^{\perp}_{\Psi_m}(\Delta S)$. Here $\Delta S$ measures the difference in the dipole moment of the positive and negative charge in an event (see Ref~\cite{Magdy:2017yje} for details). 
%
The shape of $R_{\Psi_2}(\Delta S)$ will be sensitive to CME as well as non-CME background whereas $R_{\Psi_3}(\Delta S)$ is purely driven by non-CME background and serves as a baseline.  
Model calculations have established several unique features of this observable: 1) presence of CME signal will lead to a concave shape of the $R_{\Psi_2}(\Delta S)$, 2) increasing strength of CME signal will increase the concavity of $R_{\Psi_2}(\Delta S)$, 3) in presence of CME, the concavity of $R_{\Psi_2}(\Delta S)$ will be larger than that of $R_{\Psi_3}(\Delta S)$. The measurement of $R_{\Psi_m}$ is shown in Fig.\ref{rcorrelator}. The quantity $\Delta S^{\prime\prime}$ shown is a slight variant of $(\Delta S)$ that incorporates correction for particle number fluctuations and event plane resolution. The observation of Fig.\ref{rcorrelator} indicates more concave shape for $R_{\Psi_2}$ compared to $R_{\Psi_3}$ in Au+Au whereas flat or convex shapes for p/d+Au indicates that the measurements are consistent with expectations of CME~\cite{rcorrelator}. For isobar collisions the case of CME will be confirmed if: 1) a concave shape is observed for the ratio of the observables  $R_{\Psi_2}(\Delta S)^{\rm Ru+Ru}/R_{\Psi_2}(\Delta S)^{\rm Zr+Zr}$ and 2) the concavity should be weaker for $R_{\Psi_3}(\Delta S)^{\rm Ru+Ru}/R_{\Psi_3}(\Delta S)^{\rm Zr+Zr}$.     


\section{Alternate measure: The signed Balance function}
A very recently proposed observable to search for CME is the signed balance function (SBF)~\cite{Tang:2019pbl}. The idea is to account for the ordering of the momentum of charged pairs measured by the width of SBF that is expected to be different for out-of-plane as compared to in-plane measurement captured in the ratio $r_{\rm lab}$. In addition, one can also account for the boost due to collective expansion of the system that forces all pairs to move in the same direction and measure the ratio in pair’s rest frame $r_{\rm rest}$. In presence of CME, the individual ratios as well as the double ratio $R_B=r_{\rm rest}/r_{\rm lab}$ is expected to be greater than unity. The preliminary measurements shown in Fig.\ref{rcorrelator} (right) from STAR in Au+Au 200 GeV seem to be consistent with CME expectation. This observable will be studied with the isobar data in STAR but not as a part of the blind analysis and the CME expectation will be: 1) $r({\rm Ru+Ru})>r({\rm Zr+Zr})$, and 2) $R_B({\rm Ru+Ru})>R_B({\rm Zr+Zr})$. 

%
\section{Steps for blind analysis of the isobar data from STAR}

\subsection{Modality of isobar running at RHIC}
It is better to start with a short background on the activities that preceded the isobar blind analysis in STAR.  
The idea of colliding isobar, particularly Ru+Ru and Zr+Zr to make a decisive test of CME was proposed by Voloshin in Ref~\cite{Voloshin:2010ut}, the same paper which also proposed to use Uranium collisions to disentangle signal and background of CME. The possible difference in the signals relies on $10-18\%$ higher B-field in Ru+Ru compared to Zr+Zr~\cite{Deng:2018dut} in contrast to about $4\%$ difference in flow driven background~\cite{Schenke:2019ruo}. 
Such estimates are sensitive to details of shapes, charge distribution and neutron skin thickness of the two isobar nuclei~\cite{Deng:2018dut,Xu:2017zcn,Hammelmann:2019vwd}. In the 2017-18 RHIC beam user request~\cite{starbur17} STAR collaboration therefore proposed to collect data for two 3.5 week runs in the year 2018. The projection was based on the prospect of achieving five-sigma significance or better in a scenario where the measurement of $\Delta\gamma$ has $80\%$ non-CME background. This however corresponds to the fact that the systematic uncertainty in the measurements has to be within a few percent and below the statistical significance of the measurements, something that has never been attempted before in the correlation measurements from STAR. 
This started a large scale collaboration wide effort in synergy with the RHIC collider accelerator department to plan for the isobar running in the year 2018. Based on the studies of previous years of data from Au+Au and U+U collisions several major sources of systematics in the measurement of $\Delta\gamma$ were identified. The major sources include: run-to-run variation of detector response due to loss of acceptance, change in efficiency and variation in luminosity that affects the number of reconstructed tracks in the Time Projection Chamber. This eventually leads to uncorrectable systematic uncertainties in $\Delta\gamma$. In order to minimize such systematics the proposal were to: 1) switch species in RHIC between stores e.g., in orders like Ru+Ru, Zr+Zr, Ru+Ru and so on and, 2) keep long stores to level the luminosity aiming for specific rates in the coincidence measurements of beam fragments by the STAR zero-degree calorimeters. The aim was to maintain exact balance of run and detector conditions for the two species so that observations in the two systems are equally affected and can later on be largely eliminated in the ratios of observables. 

\subsection{Blinding of data sets and preparation for analysis}
With the successful conclusion of the isobar run in the year 2018 STAR experiment collected more than 3 billion events for each isobar species. The next step was to develop the plans for a blind analysis, the main idea behind which is to eliminate predetermined biases. A total of five institutional groups are expected to perform the analysis of the data. The analysts from each group will focus on a specific aspect of the analysis described in the previous section although in many cases there are substantial overlap in some analyses that will help cross check the results. 
An important part of the blind analysis is the blinding of the data. 
The details of the blinding of the data structure is decided by members of a blinding-committee who are not part of the team of analysts and will work in close collaboration with STAR experts who are part of the production team. The idea is to provide the analysts the access to data in files where species-specific information are disguised or removed before the final step of unblinding. A careful consideration is taken by the blinding-committee to make sure the essential information available to do the analysis specific quality assurance of the data by the analysts. Some of the quality assurance, calibration and centrality determination that require species information are done only by STAR experts who are not a part of the team of analysts. Above all, the main goal of the committee is to make sure that under no circumstances physics analysts can access un-blinded data that can jeopardize the blind analysis. For example, all the data sets are produced with pseudo-run-number that cannot be used by the analysts to retrieve the exact species information. 

\subsection{Methods for the isobar blind analysis}
\begin{figure}
\includegraphics[width=1.\textwidth]{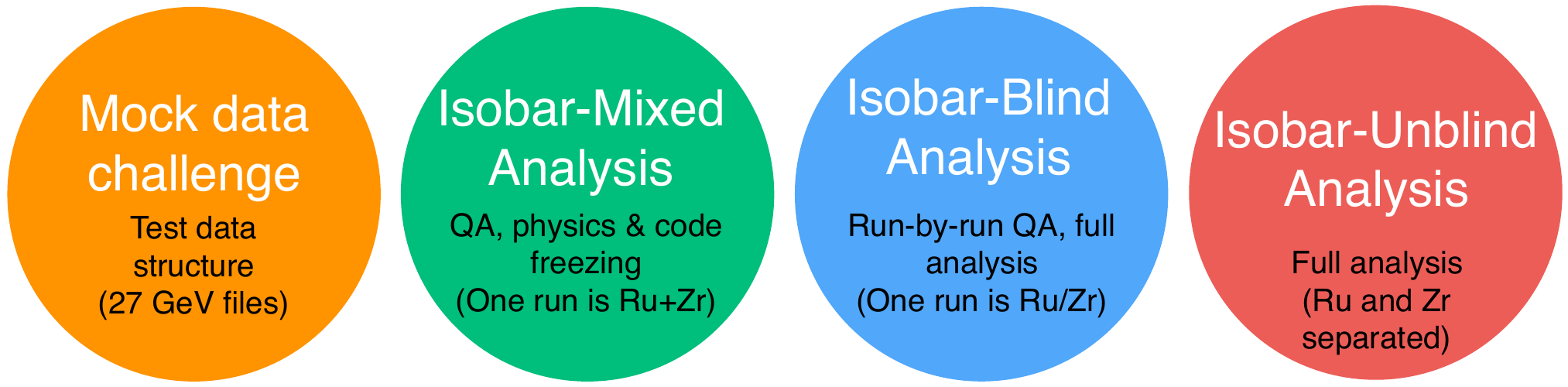}
\caption{\label{blindanalysis} The steps of isobar blind analysis. This cartoon is based on the procedure for the blind analysis of isobar data that have been outlined in Ref~\cite{Adam:2019fbq}.}
\end{figure}

The detailed procedure for the blind analysis of isobar data have been outlined in Ref~\cite{Adam:2019fbq}. Figure.\ref{blindanalysis} is a cartoon that summarizes the four steps and the main idea.
%
%

In the zeroth step shown in (by orange circle) the extreme left of Fig.\ref{blindanalysis} is the mock data challenge which is not exactly a step of the isobar data analysis but a crucial step to familiarize the analysts with the technicalities of the data structures that have been specifically designed for blind analysis. 

 The first step shown in Fig.\ref{blindanalysis} (by green circle) as the ``isobar-mixed analysis" or ``mixed-blind analysis" is truly the first step of blind analysis. This is also the most challenging steps from the point of view of the analysts. In this step the analysts are provided with data sample where each run comprise of events that are “mix” samples from two species. In this step the analysts perform the full quality assurance (QA) and physics analysis of the data, document every details of steps of the procedure and freeze the codes. After the completion of this step no changes to the analysis code is permissible. Also, no changes in the analysis procedure is allowed. The only permissible change in the following step is to reject bad runs or pile-up events. However, in order to avoid predetermined bias in analysis such rejection cannot be done arbitrarily and an automated algorithm must be developed in this step and the related codes have to be frozen. The stability of the automated QA algorithm is tested with some of the existing data sets of Au+Au and U+U collisions. 

 The second step shown in Fig.\ref{blindanalysis} (by blue circle) is referred to as the ``isobar-blind analysis" or ``unmixed-blind analysis". From this step on-wards the analysts are allowed to run their previously frozen codes. The main purpose of this step is to perform run-by-run QA of the data sample. For this the analysts are provided with files each of which contain data from a single species that is either Ru or Zr. However, there are two conditions: the files contain limited number of events that cannot lead to any statistically significant result and the species information is not revealed. Although a pseudo-run-number is used for each file, the time ordering is preserved with a unique mapping that is unknown to the analysts. It is important to maintain the time ordering to identify time-dependent changes in detectors and run conditions as a part of the run-by-run quality assurance. With this limited data sample the analysts need run the frozen automated algorithm to identify bad runs. A similar automated algorithm is also used for identifying and rejecting bad runs. After this step no more changes are allowed in terms of QA. 


The final step of isobar blind analysis is shown by red circle in Fig.\ref{blindanalysis} is referred to as ``isobar-unblind" analysis. In this step the species information will be revealed and the physics results will be produced by the analysts using the previously frozen codes. The finding from this step will be directly submitted for publication without any kind of alteration. If a mistake is found in the analysis code, the erroneous results will also accompany the corrected results. 

\section{Post-isobar era and prospects for CME search at lower collision energies}
\begin{figure}
    \centering
    \includegraphics[width=1.\textwidth]{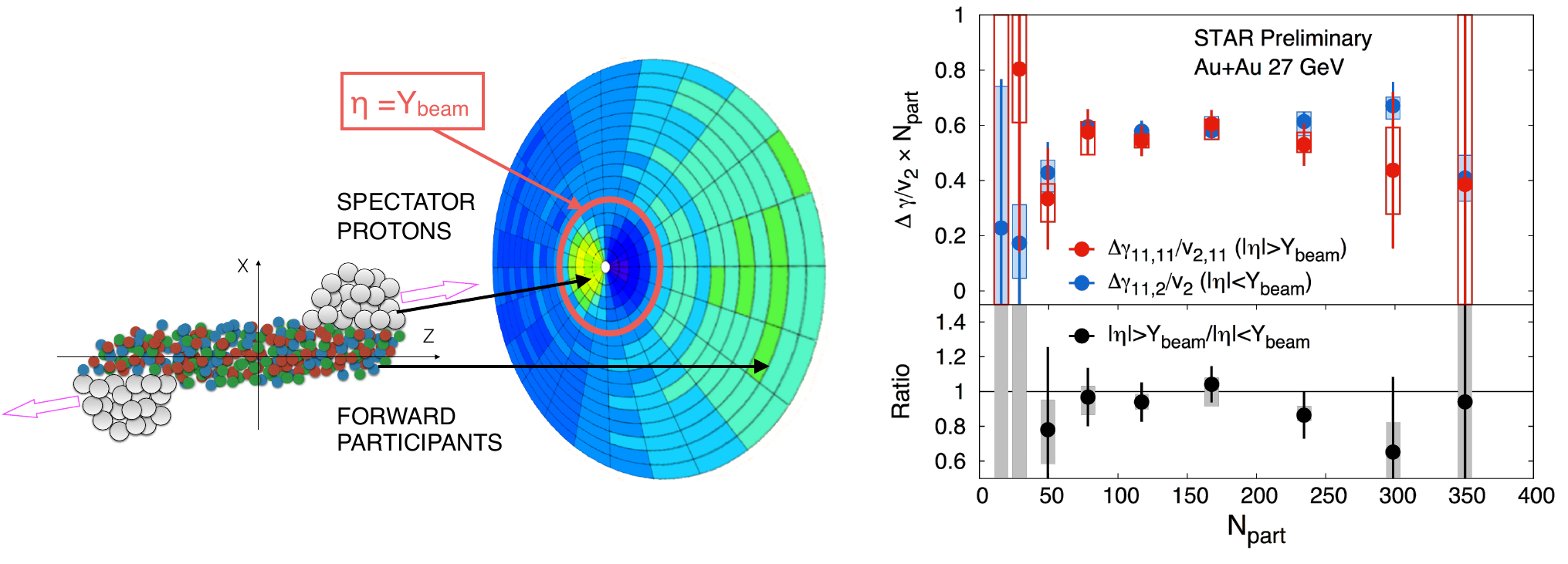}
    \caption{(Left) Figure showing EPD detector acceptance cover beam rapidity and detecting both forward participants and spectators in 27 GeV Au+Au collisions. (Right)  $\gamma$-correlators scaled by $v_2$ across different event-planes and double ratio of spectators/participant event planes which should be unity for no-CME scenario.}
    \label{fig_epd_cme}
\end{figure}

Regardless of the outcome of the measurements with the isobar program, that will be performed at the top RHIC energy, one question will remain~\cite{starbur20}. What happens at lower collision energy? In this context a new idea has emerged. The newly installed event-plane detector (EPD) upgrade provides a new capability at STAR towards CME search at lower collision energy and for the Beam Energy Scan phase-II program~\cite{Adams:2019fpo}. The idea is simple, at lower energies EPD acceptance ($2.1<|\eta|<5.1$) falls in the region of beam rapidity ($Y_{\rm beam}$) and can measure the plane of strong directed flow ($\Psi_1$) of spectator protons, beam fragments and stopped protons, therefore strongly correlated to the B-field direction (See fig\ref{fig_epd_cme}). The next step is to measure $\Delta\gamma$ with respect to $\Psi_1$ and compare it with the measurement of $\Delta\gamma$ along $\Psi_2$ planes from outer regions of EPD and TPC at mid-rapidity  that are weakly correlated to the B-field directions. A test of CME scenario will be to see if large difference is observed in the measurements. First preliminary measurements from STAR as shown in Fig~\ref{fig_epd_cme} is dominated by uncertainty but seems to show a lot of prospects for the CME search at lower energies.

\section{Summary}
Despite several challenges experimental efforts have been continued towards disentangling the CME signals from non-CME background in the measurement of charge separation across reaction plane. 
The highly anticipated results from the blind analysis of isobar collisions data provides us the best opportunity to make a decisive test of the CME in heavy-ion  collisions.

 \section{Reference}
\bibliographystyle{iopart-num}
\bibliography{spires}

\end{document}